\documentclass{article}
\usepackage[a4paper,margin=1in]{geometry}
\usepackage{amsmath,amssymb}
\usepackage{graphicx}
\usepackage{bm}
\usepackage{siunitx}
\usepackage{hyperref}
\usepackage{xcolor}
\usepackage{cite}
\usepackage{setspace}

\title{\bf Enhanced and Tunable Superconductivity Enabled by Mechanically Stable Halogen-Functionalized Mo$_2$C MXenes}

\author{
Jakkapat Seeyangnok$^{1a}$,
Udomsilp Pinsook$^{1b}$\\[1ex]
$^{1}$Department of Physics, Faculty of Science, \\
Chulalongkorn University, Bangkok, Thailand\\[1ex]
\texttt{$^{a}$Jakkapat.Se@chula.ac.th}\\
\texttt{$^{b}$Udomsilp.P@chula.ac.th}\\
}

\date{} % no date

\begin{document}

\maketitle

% =========================
% Abstract
% =========================
\begin{abstract}
We present a comprehensive first-principles investigation of the structural, electronic, vibrational, and superconducting properties of halogen-functionalized Mo$_2$Y$X_2$ ($Y$= C, N;$X=$ F, Cl, Br, I) MXene monolayers. Density functional theory and density functional perturbation theory calculations reveal that, among the halogenated systems considered, only Br- and I-functionalized Mo$_2$C monolayers are dynamically stable, as confirmed by positive-definite phonon spectra throughout the Brillouin zone. Electronic structure calculations show metallic behavior with states near the Fermi level dominated by Mo $d$ orbitals with pronouced electronic density of states providing favorable conditions for strong electron–phonon coupling (EPC). The resulting EPC constants place both systems in the strong-coupling regime, yielding superconducting transition temperatures of $T_c = 13.1$ K for Mo$_2$CBr$_2$ and $T_c = 18.1$ K for Mo$_2$CI$_2$ within the Allen–Dynes formalism. Notably, halogen functionalization itself plays a crucial role in enhancing superconductivity in Mo$_2$C of $T_c = 7.2$ K, leading to a substantial increase in the superconducting transition temperature compared with pristine Mo$_2$C through strengthened electron–phonon coupling. Furthermore, we demonstrate that superconductivity in these systems is highly tunable via carrier doping and biaxial tensile strain. Electron doping significantly enhances EPC and raises $T_c$ up to 21.7 K for Mo$_2$CBr$_2$ and 21.3 K for Mo$_2$CI$_2$. Our results identify halogen-functionalized Mo$_2$C MXenes as mechanically robust, phonon-mediated two-dimensional superconductors and highlight carrier doping as an effective strategy for optimizing their superconducting performance.
\end{abstract}

% =========================
% Keywords
% =========================

% =========================
\section{Introduction}
% =========================
Achieving superconductivity with high transition temperatures together with robust structural stability remains a central objective in condensed matter physics~\cite{frohlich1950theory,migdal1958interaction,eliashberg1960interactions,nambu1960quasi}. Hydrogen-rich compounds have demonstrated remarkable progress in this direction, exhibiting record superconducting transition temperatures exceeding 200~K under extreme compression~\cite{duan2014pressure,peng2017hydrogen,liu2017potential}. However, the requirement of megabar pressures severely limits their practical applicability~\cite{drozdov2015conventional,einaga2016crystal,drozdov2019superconductivity,somayazulu2019evidence}. These breakthroughs originate from Ashcroft’s seminal proposal that strong electron--phonon coupling (EPC) in metallic hydrogen and hydrogen-dominated materials can support high-$T_c$ superconductivity~\cite{ashcroft1968metallic,ashcroft2004hydrogen}, motivating the search for alternative phonon-mediated superconductors that are stable at ambient conditions and whose properties can be systematically engineered.

Two-dimensional (2D) materials have emerged as a versatile platform for exploring reduced-dimensionality physics, tunable quantum states, and novel electronic phenomena. The convergence of these developments has driven growing interest in combining the microscopic design principles of conventional phonon-mediated superconductivity with the intrinsic tunability of 2D materials, aiming to realize superconductivity at reduced or ambient pressures. Hydrogen-rich 2D systems have therefore attracted considerable attention as promising platforms for phonon-mediated superconductivity, as the light mass of hydrogen can significantly enhance EPC while preserving structural flexibility. Early theoretical studies proposed that fully hydrogenated graphene (graphane) could host superconductivity with critical temperatures exceeding 90~K~\cite{sofo2007graphane}. This idea was subsequently extended to a wide range of ternary and multicomponent 2D hydrides, including hydrogen-functionalized MgB$_2$ monolayers with predicted $T_c$ values around 67~K~\cite{savini2010first}, and numerous other hydrogenated 2D systems~\cite{li2022phonon,bekaert2019hydrogen,jsee_2dm2x_jpcs,jsee_GaXS2_jap,han2023high,liu2024three,seeyangnok2025high_npj2d,seeyangnok2025hydrogenation_nanoscale,jsee_mb4h_arxiv}. Collectively, these studies establish chemical functionalization as an effective route for inducing or enhancing superconductivity in low-dimensional materials. A representative example is the experimentally realized Janus 2H-MoSH monolayer~\cite{lu2017janus}, which was theoretically predicted to exhibit phonon-mediated superconductivity with a critical temperature of approximately 26.8~K~\cite{liu2022two,ku2023ab}, with extensions to bilayer systems~\cite{pinsook2025superconductivity} with possible charge density wave~\cite{seeyangnok2026charge} including Li functionalization~\cite{xie2024strong,moseliseeyangnok}. The WSH and WSeH were shown to be dynamically stable superconductors with $T_c$ values exceeding 12~K~\cite{wseh_prb,wsh_2dmat}, results later confirmed independently~\cite{gan2024hydrogenation,fu2024superconductivity}. These advances highlight a broader design principle: surface functionalization can stabilize low-dimensional lattices while simultaneously enhancing phonon-mediated superconductivity. Similar concepts have been explored in group-IV transition-metal dichalcogenides~\cite{joseph2023review,lasek2021synthesis,mattinen2019atomic,zhang2016systematic,toh2016catalytic,xie2015two}, where functionalization-induced superconductivity has been theoretically predicted in several Janus transition-metal chalcogenide hydrides with transition temperatures in the 10–30~K range~\cite{li2024machine,ul2024superconductivity} with competing magnetic ground states~\cite{seeyangnok2025competition}, analogous to those observed in C$_6$H~\cite{jsee_c6h_jccm} and CrSH~\cite{sukserm2025half}. This interplay between superconductivity, magnetism, and structural stability underscores the importance of identifying functionalization strategies that yield both robust lattices and strong EPC.

MXenes have emerged as a highly promising materials platform for investigating superconductivity driven by surface functionalization. Several theoretical studies have explored the influence of hydrogen incorporation on their superconducting properties. For example, hydrogenated Mo$_2$C has been predicted to exhibit superconductivity with a transition temperature of approximately 13~K~\cite{lei2017predicting}. In Ti-based MXenes~\cite{tsuppayakorn2023hydrogen}, hydrogenation induces superconducting behavior in Ti$_2$CH, Ti$_2$CH$_2$, and Ti$_2$CH$_4$, with reported transition temperatures of 0.2~K, 2.3~K, and 9.0~K, respectively, while Ti$_2$CSH is predicted to host a notably higher $T_c$ of 22.6~K~\cite{seeyangnok2025theoretical_Ti2CSH}. Mo-based MXenes have also been extensively studied owing to their rich and tunable physical properties, particularly the superconducting behavior observed in Mo$_2$C and Mo$_2$N~\cite{bekaert2020first,pereira2022strain,liu2023theoretical}, as well as the enhancement of superconductivity induced by hydrogenation~\cite{bekaert2022enhancing}. For Mo$_2$C, superconductivity has been predicted in both the 2H and 1T polymorphs, with calculated transition temperatures of 7.1~K~\cite{bekaert2020first} and 3.2~K~\cite{lei2017predicting}, respectively. In comparison, Mo$_2$N exhibits stronger superconductivity, with an estimated $T_c$ of 16.0~K in the relaxed structure~\cite{bekaert2020first}, which can be further enhanced to 24.7~K under applied strain~\cite{pereira2022strain}. Beyond Mo-based systems, V- and Zr-based MXenes, including V$_2$C, V$_2$N, Zr$_2$C, and Zr$_2$N, have been widely investigated for diverse functionalities, ranging from optical properties~\cite{lee2021investigation} and hydrogen storage capabilities~\cite{saharan2024v2n,yorulmaz2020systematical} to applications in electrochemical energy storage devices~\cite{liu2022two,meng2018theoretical,behl2024recent}. Recently, a functionalized Nb$_2$CCl$_2$ has also been reported to exhibit enhanced superconductivity~\cite{sevik2023superconductivity}. Taken together, these studies underscore the versatility of MXenes and highlight their potential as a flexible platform for extending functionalization-driven superconductivity beyond hydrogenated systems. 

In this work, we perform a comprehensive first-principles investigation of halogen-functionalized Mo$_2$Y$X_2$ MXene monolayers ($Y=$ C, N; $X=$ F, Cl, Br, I), focusing on their structural and mechanical stability, electronic structure, lattice dynamics, and phonon-mediated superconductivity. Among the systems considered, only Mo$_2$C$X_2$ ($X=$ Br, I) are found to be dynamically stable. Using density functional theory and density functional perturbation theory, we demonstrate that Br- and I-functionalized Mo$_2$C monolayers are both dynamically and mechanically stable, exhibit strong electron--phonon coupling, and support phonon-mediated superconductivity. Furthermore, we investigate the tunability of superconductivity under carrier doping and biaxial tensile strain, identifying electron doping as an efficient strategy for enhancing the superconducting transition temperature.

% =========================
\section{Computational Methods}
All first-principles calculations were performed within density functional theory (DFT) as implemented in the \textsc{Quantum ESPRESSO} package~\cite{giannozzi2009quantum, giannozzi2017advanced}. The electronic wave functions were expanded in a plane-wave basis set with a kinetic-energy cutoff of 80~Ry, while a charge-density cutoff of 240~Ry was employed. Brillouin-zone integrations were carried out using a $24 \times 24 \times 1$ Monkhorst--Pack $k$-point mesh~\cite{monkhorst1976special}, with van der Waals interactions taken into account. Structural relaxations were performed until the residual forces on each atom were less than $10^{-5}$~Ry/Bohr. The generalized gradient approximation (GGA) with the Perdew–Burke–Ernzerhof (PBE) functional~\cite{perdew1996generalized} with Norm-conserving optimized Vanderbilt (ONCV) pseudopotentials~\cite{hamann2013optimized, schlipf2015optimization} were used to describe the electron--ion interactions. Metallic occupations were treated using Methfessel--Paxton smearing with a smearing width of 0.02~Ry~\cite{marzari1999thermal}. 

The electron--phonon coupling (EPC) calculations were performed using density functional perturbation theory (DFPT) \cite{baroni2001phonons}. The phonon wave vectors were sampled on a $12 \times 12 \times 1$ $q$-point mesh. The superconducting transition temperature $T_c$ was estimated using the isotropic Eliashberg spectral function $\alpha^2F(\omega)$ within the Allen--Dynes formalism~~\cite{allen1975transition,pinsook2024analytic} given by
\begin{equation}
T_c = \frac{\omega_{\ln}}{1.2}
\exp\left[
-\frac{1.04(1+\lambda)}
{\lambda - \mu^{\ast}(1+0.62\lambda)}
\right],
\end{equation}
where $\lambda$ is the electron--phonon coupling constant and $\mu^{\ast}$ is the Coulomb pseudopotential.

The EPC constant $\lambda$ is obtained from the Eliashberg spectral function as
\begin{equation}
\lambda = 2 \int_0^{\infty} \frac{\alpha^2F(\omega)}{\omega} \, d\omega.
\end{equation}

The logarithmic average phonon frequency $\omega_{\ln}$ is defined as
\begin{equation}
\omega_{\ln} =
\exp\left[
\frac{2}{\lambda}
\int_0^{\infty}
\frac{\alpha^2F(\omega)}{\omega}
\ln \omega \, d\omega
\right],
\end{equation}
while the second moment of the phonon spectrum $\omega_2$ is given by
\begin{equation}
\omega_2 =
\left[
\frac{2}{\lambda}
\int_0^{\infty}
\alpha^2F(\omega)\,\omega \, d\omega
\right]^{1/2}.
\end{equation}

\section{Results and Discussion}
\subsection{Structural Properties}
%\subsection{Crystal Structures}
\begin{figure}[h!]
\centering
\includegraphics[width=10cm]{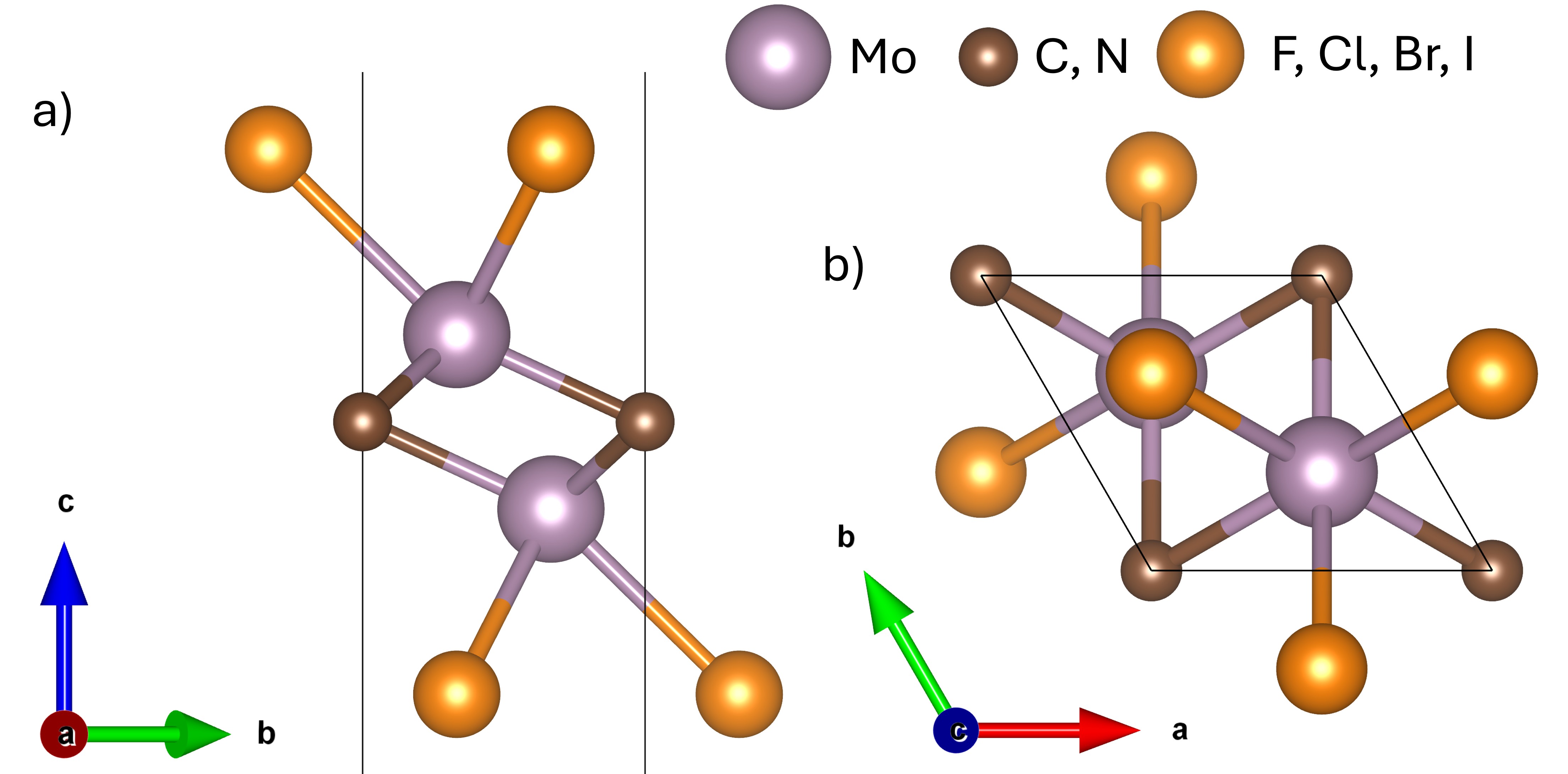}
\caption{Crystal structure of halogen-functionalized Mo$_2$C and Mo$_2$N MXenes, Mo$_2$C$X_2$ ($X=$ F, Cl, Br, I). (a) Side view and (b) top view of the optimized monolayer structure. Purple, brown, and orange spheres represent Mo, C (or N), and halogen atoms, respectively.}
\label{fig:structure}
\end{figure}

The primitive cell of the halogen-functionalized Mo$_2$C and Mo$_2$N monolayer crystallizes in a hexagonal Bravais lattice of the trigonal space group $P\bar{3}m1$ (No.~164) as shown in Figure~\ref{fig:structure}.  Within the primitive cell, the carbon atom occupies the high-symmetry position at $(0,0,0.5)$, while the two Mo atoms are located at fractional coordinates $(2/3,1/3)$ and $(1/3,2/3)$, consistent with the centrosymmetric nature of the lattice. The halogen atoms ($X=$ F, Cl, Br, I) are adsorbed on both sides of the Mo$_2$C layer and occupy symmetry-equivalent Wyckoff positions characterized by in-plane fractional coordinates $(1/3,2/3)$ and $(2/3,1/3)$, while their out-of-plane coordinates are fully optimized during structural relaxation. 

%\subsection{Structural Stability}
The dynamical stability of the considered structures was first examined by calculating their phonon spectra to ensure the absence of lattice instabilities. Among the investigated systems, only Br- and I-functionalized Mo$_2$C monolayers exhibit dynamical stability, as evidenced by positive-definite phonon frequencies throughout the Brillouin zone, as discussed in Section~\ref{phononsection}. In this subsection, we further investigate the thermal and mechanical stability of the Br- and I-functionalized Mo$_2$C monolayers. In addition, various magnetic polarization configurations, including ferromagnetic (FM), G-type antiferromagnetic (GAF), and A-type antiferromagnetic (AAF) orderings, were examined, with the nonmagnetic (NM) state identified as the ground state. The optimized lattice constants are 3.40~\AA\ and 3.52~\AA\ for the Br- and I-functionalized Mo$_2$C monolayers, respectively. These systems exhibit negative binding energies of $-5.51$~eV and $-4.26$~eV for Mo$_2$C$X_2$ ($X=$ Br, I), respectively, indicating their energetic stability. The binding energy $E_{\mathrm{b}}$ is defined as
\begin{equation}
E_{\mathrm{b}} =
E_{\mathrm{Mo_2C}X_2}
-
E_{\mathrm{Mo_2C}}
-
E_{X_{2}},
\qquad X = \mathrm{Br},\,\mathrm{I},
\end{equation}
where $E_{\mathrm{Mo_2C}X_2}$ is the total energy of the halogen-functionalized Mo$_2$C monolayer, $E_{\mathrm{Mo_2C}}$ is the total energy of the pristine Mo$_2$C monolayer, and $E_{X_2}$ is the total energy of the halogen molecule ($X_2 =$ Br$_2$ or I$_2$).

To evaluate the mechanical stability of Mo$_2$C$X_2$ ($X=$ Br, I) monolayers, we calculated the in-plane elastic constants using strain--energy relationships appropriate for two-dimensional hexagonal systems. The elastic constants $C_{ij}$ were obtained from the second derivatives of the total energy with respect to the applied in-plane strain components according to
\begin{equation}
C_{ij} = \frac{1}{S_0} \frac{\partial^2 E}{\partial \epsilon_i \partial \epsilon_j},
\end{equation}
where $S_0$ is the equilibrium area of the unit cell, and $\epsilon_i$ and $\epsilon_j$ denote the in-plane strain components.

For hexagonal lattices, symmetry reduces the number of independent in-plane elastic constants to two, namely $C_{11}$ and $C_{12}$, with $C_{11} = C_{22}$. The shear modulus is defined as $C_{66} = (C_{11} - C_{12})/2$. These elastic constants characterize the linear stress--strain response of the monolayers and can be expressed through the generalized Hooke’s law for two-dimensional materials as
\begin{equation}
\sigma =
\begin{bmatrix}
C_{11} & C_{12} & 0 \\
C_{12} & C_{11} & 0 \\
0 & 0 & C_{66}
\end{bmatrix}
\varepsilon.
\end{equation}

\begin{table}[h!]
\centering
\caption{In-plane elastic constants ($C_{11}$, $C_{22}$), off-diagonal elastic constant ($C_{12}$), and shear modulus ($C_{66}$) for Mo$_2$C$X_2$ ($X=$ Br, I) monolayers. All values are given in eV/\AA$^{2}$.}
\begin{tabular}{|c|c|c|c|}
\hline
2D compound & $C_{11}$, $C_{22}$ & $C_{12}$ & $C_{66}$ \\
\hline
Mo$_2$CBr$_2$ & 13.44 & 4.90 & 4.27 \\
Mo$_2$CI$_2$  & 10.71 & 3.88 & 3.42 \\
\hline
\end{tabular}
\label{tab:elasticity}
\end{table}

The calculated elastic constants, summarized in Table~\ref{tab:elasticity}, satisfy the mechanical stability criteria for two-dimensional systems, namely $C_{11}C_{22} - C_{12}^2 > 0$ and $C_{11}, C_{22}, C_{66} > 0$, in accordance with the conditions proposed by Mouhat and Coudert~\cite{mouhat2014necessary}. These results confirm the mechanical stability of the Br- and I-functionalized Mo$_2$C monolayers.

\subsection{Electronic Properties}
\begin{figure}[h!]
\centering
\includegraphics[width=16cm]{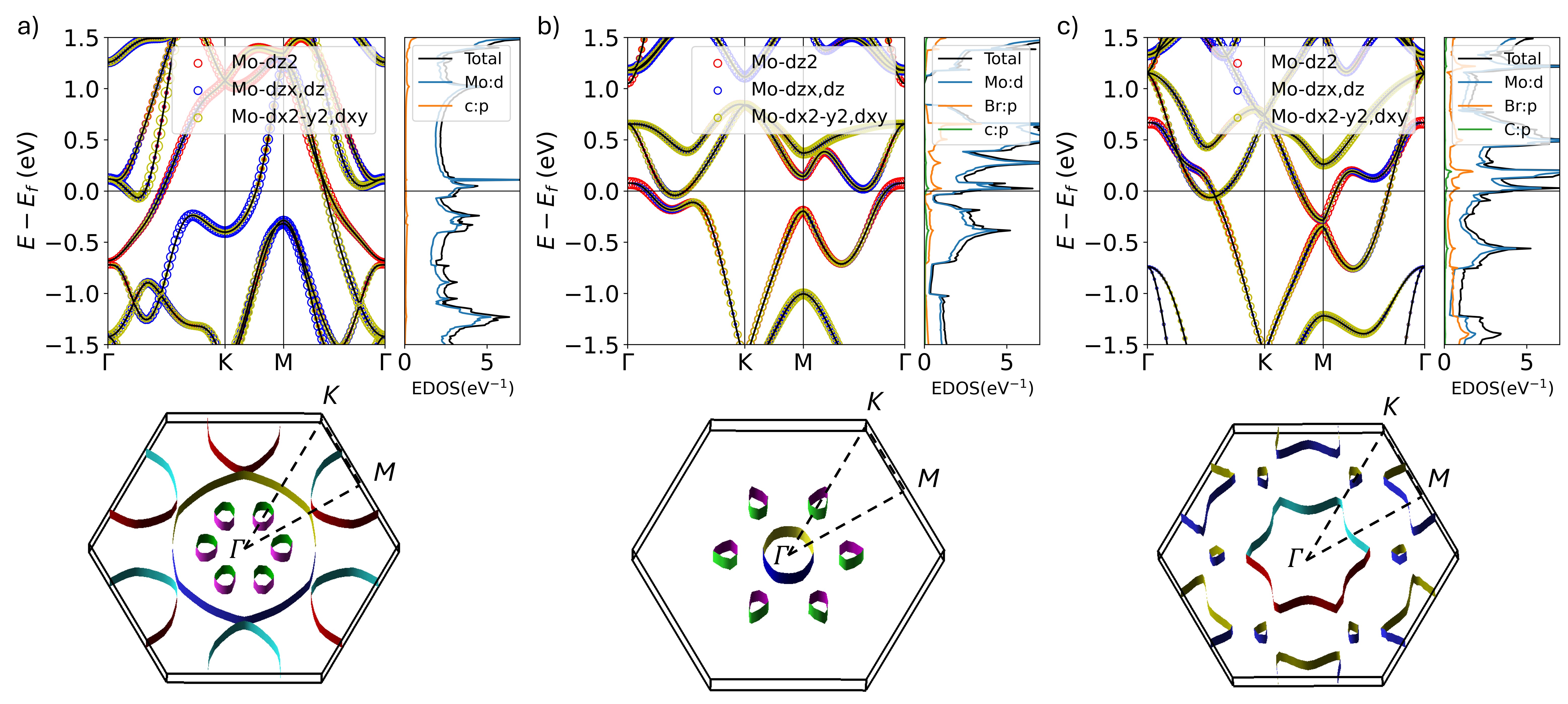}
\caption{Electronic structure of halogen-functionalized Mo$_2$C$X_2$ monolayers.
(a)–(c) Orbital-projected electronic band structures and the corresponding projected density of states (PDOS) of pristine Mo$2$C, Mo$2$Br$2$, and Mo$2$CI$2$, together with their respective Fermi surfaces. Colored circles denote the orbital contributions from Mo $d{z^2}$, Mo $d{zx}/d{zy}$, and Mo $d{x^2-y^2}/d{xy}$ states, while the PDOS highlights the contributions from Mo-$d$, halogen-$p$, and C-$p$ orbitals.}
\label{fig:band_fermi}
\end{figure}

The electronic properties of pristine, Br- and I-functionalized Mo$_2$C monolayers were investigated using orbital-projected band structure calculations, projected density of states (PDOS), and their corresponding Fermi surfaces. Figures~\ref{fig:band_fermi}(a), Figures~\ref{fig:band_fermi}(b) and \ref{fig:band_fermi}(c) present the electronic band structures together with the corresponding PDOS for Mo$_2$C, Mo$_2$CBr$_2$ and Mo$_2$CI$_2$, respectively.

These systems exhibit clear metallic behavior, characterized by multiple bands crossing the Fermi level along the high-symmetry path $\Gamma$--$K$--$M$--$\Gamma$. The PDOS analysis shows that the electronic states near the Fermi level are predominantly derived from Mo $d$ orbitals, indicating that the metallicity is primarily governed by the transition-metal sublattice. In particular, the Mo $d_{z^2}$, $d_{zx}/d_{zy}$, and $d_{x^2-y^2}/d_{xy}$ orbitals make the dominant contributions to the density of states at the Fermi level, whereas the halogen $p$ and C $p$ states contribute only marginally in this energy region.

Notably, the total electronic density of states of pristine Mo$_2$C, Mo$_2$CBr$_2$ and Mo$_2$CI$_2$ exhibits a pronounced peak near the Fermi level, indicative of a van Hove singularity. Such an enhanced density of states can be beneficial for phonon-mediated superconductivity by strengthening Cooper pairing. The calculated Fermi surfaces, shown in the lower panels of Fig.~\ref{fig:band_fermi}, further corroborate the metallic character of both monolayers. Mo$_2$CBr$_2$ displays relatively small and well-separated electron pockets centered around the $\Gamma$ point, while Mo$_2$CI$_2$ exhibits more extended and interconnected Fermi surface sheets, reflecting stronger band dispersion and a higher density of states at the Fermi level.

Therefore, the metallic electronic structures and Mo-$d$-dominated states at the Fermi level provide a favorable electronic environment for strong electron--phonon coupling, which is a key prerequisite for the emergence of superconductivity in these systems, as discussed in the following section.

%%%%%%%%%%%%%%%%%%%%%%%%%%%%%%%%%%%%%%%%%%%%%%%%%%%%%%%%%%%%%%%%%%%%%%%%%%%%%%%%%%%%%%%%%%%%%%%%%%%%%%%%%%%%%%%%%%%%%%%%%%%%%%%%%%%%%%%%%%%%%%
\subsection{Phonon and Phonon-mediated Superconductivity}\label{phononsection}
\begin{figure}[h!]
\centering
\includegraphics[width=12cm]{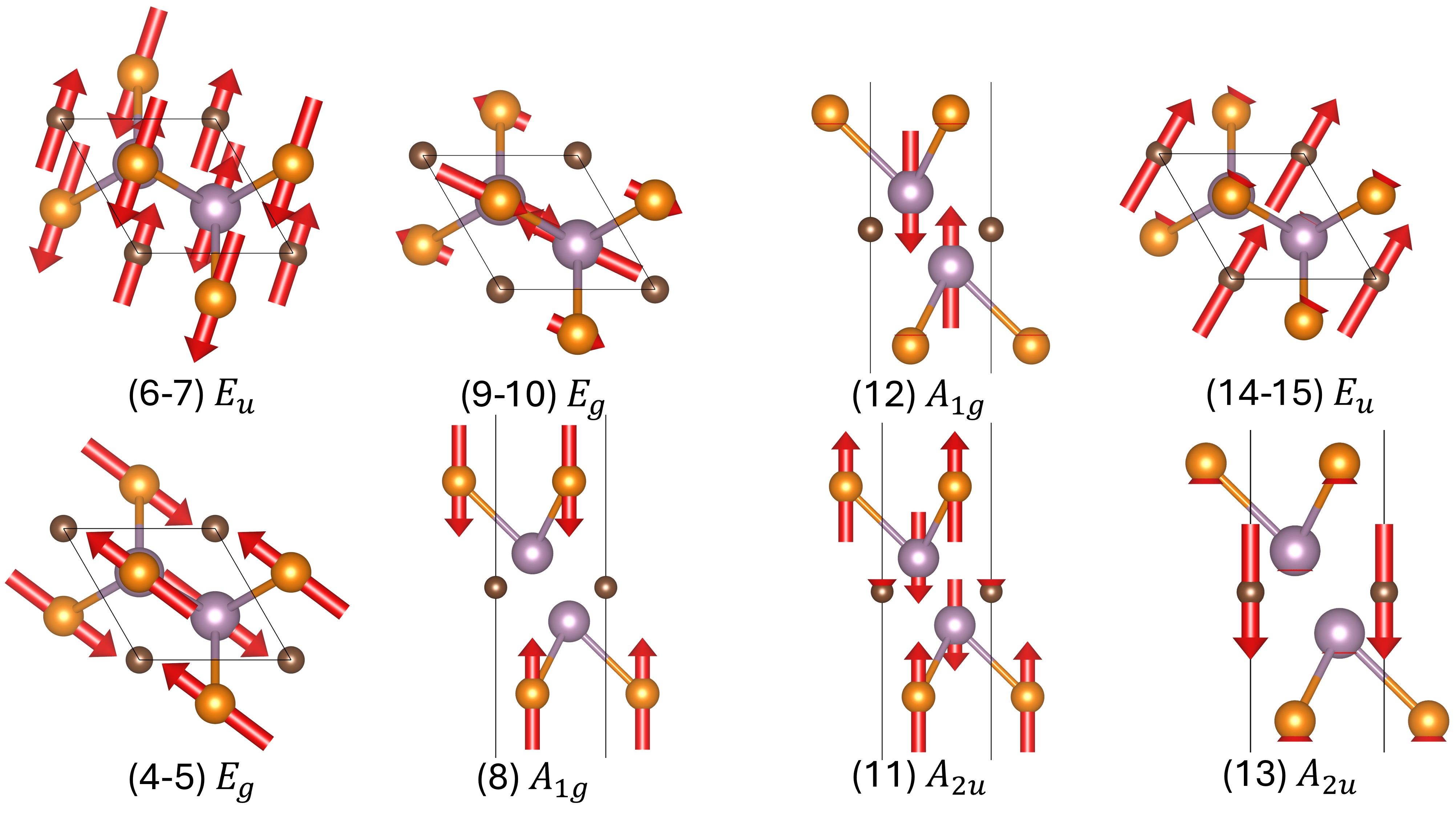}
\caption{Representative zone-center ($\Gamma$-point) phonon eigenmodes of the halogen-functionalized Mo$_2$C$X_2$ monolayer. The vibrational displacement patterns for selected Raman- and infrared-active modes are shown, including the $E_u$, $E_g$, $A_{1g}$, and $A_{2u}$ symmetries, as indicated beneath each panel together with their corresponding mode indices. Red arrows denote the atomic displacement vectors, illustrating both in-plane and out-of-plane vibrational character. Purple, brown, and orange spheres represent Mo, C, and halogen atoms, respectively.}
\label{fig:phonon_modes}
\end{figure}

\begin{figure}[h!]
    \centering
    \includegraphics[width=16cm]{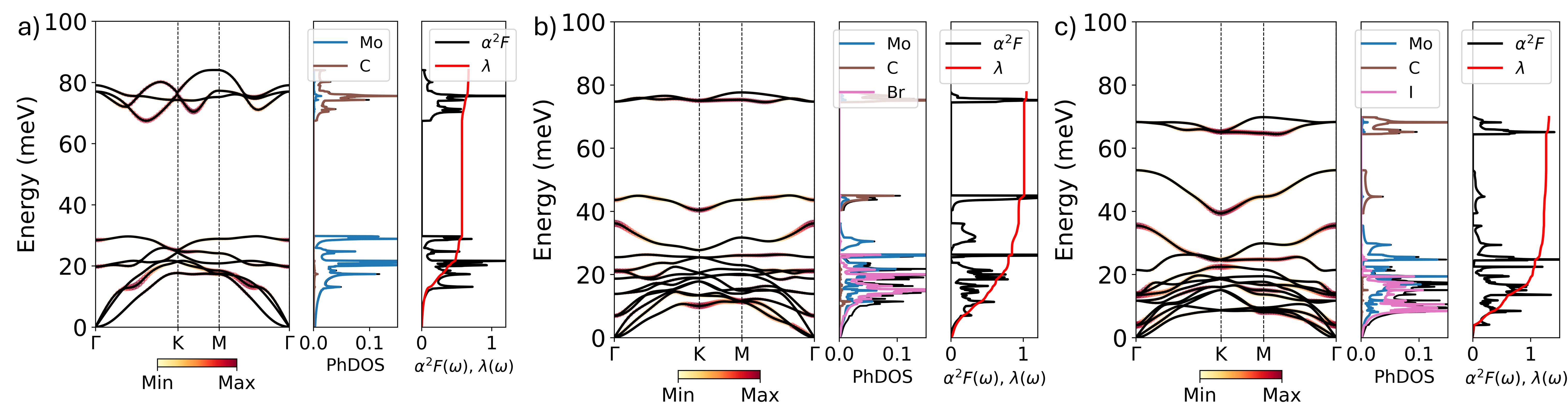}
    \caption{Phonon dispersion relations weighted by the electron--phonon coupling (EPC), phonon density of states (PhDOS), and Eliashberg spectral function of halogen-functionalized Mo$_2$C$X_2$ monolayers. (a-c) EPC-weighted phonon dispersion, atom-resolved PhDOS, Eliashberg spectral function $\alpha^2F(\omega)$, and cumulative EPC constant $\lambda(\omega)$ for Mo$_2$C, Mo$_2$CBr$_2$, and Mo$_2$CI$_2$. The phonon dispersions are plotted along the high-symmetry path $\Gamma$--$K$--$M$--$\Gamma$, with the color scale indicating the magnitude of EPC strength, from weak (blue) to strong (red). The PhDOS highlights the vibrational contributions from Mo, C, and halogen atoms, while $\alpha^2F(\omega)$ and $\lambda(\omega)$ quantify the phonon-mode-resolved EPC contributions.}
    \label{fig:phonon_epc}
    \end{figure}

The phonon properties and phonon-mediated superconductivity of Br- and I-functionalized Mo$_2$C monolayers were investigated using density functional perturbation theory. Both systems are dynamically stable, as confirmed by the absence of imaginary phonon modes throughout the Brillouin zone. Representative zone-center phonon eigenvectors corresponding to selected vibrational modes are shown in Fig.~\ref{fig:phonon_modes}. The phonon dispersions, phonon density of states (PhDOS), and electron--phonon coupling (EPC)-weighted phonon spectra are presented in Fig.~\ref{fig:phonon_epc}.

\begin{table}[h!]
\centering
\caption{Zone-center ($\Gamma$-point) optical phonon modes of Mo$_2$C$X_2$ ($X=$ Br, I). Mode symmetry, spectroscopic activity, and phonon frequencies for Br- and I-functionalized systems are listed. $R$ and $I$ denote Raman- and infrared-active modes, respectively.}
\begin{tabular}{|c|c|c|c|c|}
\hline
Symmetry & Activity & Pristine (meV) & Br (meV) & I (meV) \\
\hline
$E_g$      & Raman     & - & 6.9  & 4.2 \\
$E_u$      & Infrared  & - & 13.9 & 11.9 \\
$A_{1g}$   & Raman     & - & 18.6 & 14.0 \\
$E_g$      & Raman     & 19.8 & 21.1 & 13.6 \\
$A_{2u}$   & Infrared  & - & 25.4 & 21.5 \\
$A_{1g}$   & Raman     & 28.4 & 36.2 & 35.5 \\
$A_{2u}$   & Infrared  & 77.4 & 44.7 & 54.0 \\
$E_u$      & Infrared  & 79.8 & 75.2 & 68.7 \\
\hline
\end{tabular}
\label{tab:phonon_modes}
\end{table}

The zone-center optical phonon modes of pristine  Mo$2$C and funtionalized Mo$2$C$X_2$ ($X=$ Br, I), summarized in Table~\ref{tab:phonon_modes}, can be classified according to their irreducible representations and spectroscopic activities. Raman-active modes belong to the $E_g$ and $A{1g}$ symmetries, while infrared-active modes correspond to the $E_u$ and $A_{2u}$ representations, consistent with the $D_{3d}$ point-group symmetry of the crystal. The corresponding vibrational displacement patterns for selected modes are illustrated in Fig.~\ref{fig:phonon_modes}. The low-frequency $E_g$ modes primarily involve collective in-plane vibrations of Mo and halogen atoms, whereas higher-frequency $A_{1g}$ modes are dominated by out-of-plane symmetric motions of the halogen atoms relative to the Mo$2$C layer. Infrared-active $A{2u}$ and $E_u$ arise from relative displacements between Mo and halogen atoms along the out-of-plane and in-plane directions, respectively. These mode characteristics provide a microscopic basis for understanding the phonon contributions to electron–phonon coupling and the resulting superconducting behavior discussed below.

The calculated Eliashberg spectral functions $\alpha^2F(\omega)$ reveal that the dominant contributions to the EPC originate from low- and intermediate-frequency phonon modes, primarily associated with Mo vibrations, with additional contributions from halogen-related modes at higher frequencies. This behavior is consistent with the metallic electronic structures dominated by Mo $d$ states near the Fermi level. The cumulative EPC constant $\lambda(\omega)$ increases rapidly in the low-frequency region and gradually saturates at higher frequencies, indicating that low-energy phonons play a key role in mediating superconductivity.

For pristine Mo$2$C, the calculated electron–phonon coupling constant is $\lambda = 0.67$, with logarithmic and second moments of the phonon frequency given by $\omega{\log} = 231.6$ K and $\omega_2 = 378.0$ K, respectively. These values yield a corrected superconducting transition temperature of $T_c = 7.2$ K. For Mo$_2$CBr$_2$, the total EPC constant is found to be $\lambda = 1.04$, placing the system in the strong-coupling regime. The logarithmic average phonon frequency and second moment of the phonon spectrum are $\omega_{\ln} = 154.3$~K and $\omega_2 = 278.0$~K, respectively. Using the Allen--Dynes modified McMillan formula with a standard Coulomb pseudopotential, the superconducting transition temperature is estimated to be $T_c = 13.1$~K. In the case of Mo$_2$CI$_2$, the EPC strength is further enhanced, with a total coupling constant of $\lambda = 1.32$. Despite a slightly reduced characteristic phonon frequency ($\omega_{\ln} = 153.4$~K and $\omega_2 = 243.5$~K), the increased EPC leads to a higher superconducting transition temperature of $T_c = 18.1$~K. 

Therefore, halogen functionalization markedly enhances the superconducting properties of Mo$_2$C. In particular, Br and I functionalization substantially increase the electron–phonon coupling strength compared to pristine Mo$_2$C, driving the system from a moderate-coupling regime into a strong-coupling regime. As a result, the superconducting transition temperature is significantly elevated in Mo$_2$CBr$_2$ and Mo$_2$CI$_2$, demonstrating that halogen functionalization provides an effective route to tuning and enhancing superconductivity in Mo$_2$C-based MXene monolayers.

%%%%%%%%%%%%%%%%%%%%%%%%%%%%%%%%%%%%%%%%%%%%%%%%%%%%%%%%%%%%%%%
\subsection{Tunable Superconductivity}
%%%%%%%%%%%%%%%%%%%%%%%%%%%%%%%%%%%%%%%%%%%%%%%%%%%%%%%%%%%%%%%
    \begin{figure}[h!]
    \centering
    \includegraphics[width=7cm]{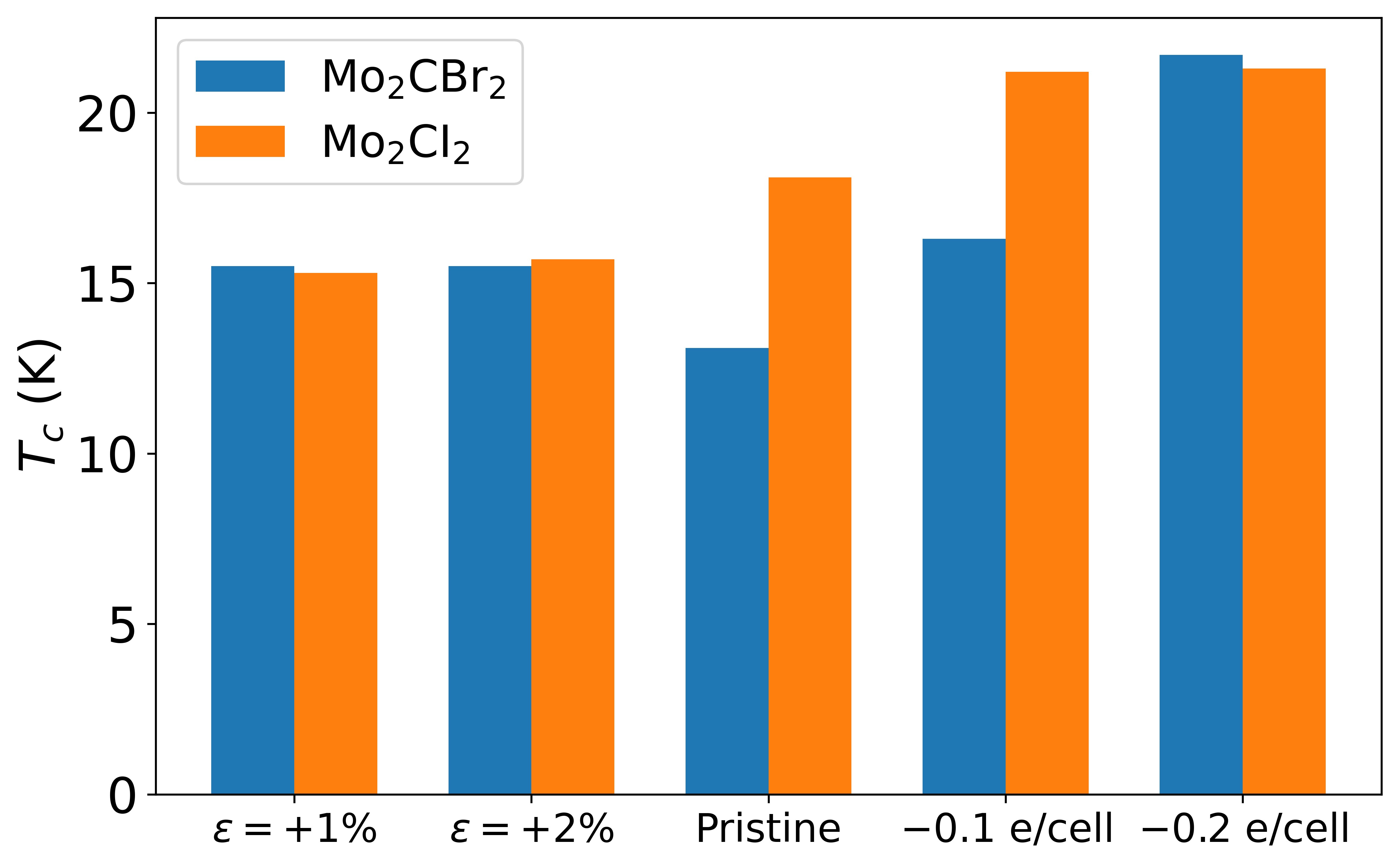}
    \caption{Tunability of the superconducting transition temperature $T_c$ of halogen-functionalized Mo$_2$C$X_2$ monolayers under biaxial tensile strain and electron doping. The calculated $T_c$ values for Mo$_2$CBr$_2$ (blue) and Mo$_2$CI$_2$ (orange) are shown for $+1\%$ and $+2\%$ biaxial strain, the pristine structures, and electron doping levels of $-0.1$ and $-0.2~e$/cell.}
    \label{fig:Tc_tunability}
    \end{figure}

The superconducting properties of Br- and I-functionalized Mo$_2$C monolayers can be effectively tuned by both carrier doping and biaxial strain. To elucidate these effects, we systematically investigated the evolution of the electron--phonon coupling (EPC) constant $\lambda$, characteristic phonon frequencies, and the resulting superconducting transition temperature $T_c$ under electron doping and tensile strain as shown in Fig.~\ref{fig:Tc_tunability}.

For Mo$_2$CBr$_2$, Upon electron doping, a pronounced enhancement of EPC is observed. At a doping level of $-0.1~e$/cell, $\lambda$ increases to 1.17, accompanied by a moderate increase in $\omega_{\ln}$, resulting in an enhanced $T_c$ of 16.3~K. Further increasing the doping concentration to $-0.2~e$/cell leads to a strong-coupling regime with $\lambda = 1.51$, despite a slight reduction in characteristic phonon frequencies, giving rise to a significantly enhanced $T_c$ of 21.7~K. Tensile strain also strongly modifies the superconducting properties of Mo$_2$CBr$_2$. Under $+1\%$ biaxial strain, the EPC constant increases to $\lambda = 1.23$, while $+1\%$ biaxial strain reduces $\omega_{\ln}$ to 139.4~K, resulting in $T_c = 15.5$~K. At $+2\%$ strain, a substantial enhancement of $\lambda$ to 1.78 is observed; however, the strong reduction in $\omega_{\ln}$ counterbalances the EPC enhancement, leading to a nearly unchanged $T_c$ of 15.5~K. This behavior highlights the competing roles of EPC strength and phonon frequency in determining $T_c$~\cite{pinsook2020search}.

A similar but more pronounced tunability is found for Mo$_2$CI$_2$. Electron doping significantly enhances EPC, with $\lambda$ increasing to 1.65 at $-0.1~e$/cell and further to 1.86 at $-0.2~e$/cell. Although phonon softening reduces $\omega_{\ln}$, the enhanced EPC leads to elevated superconducting transition temperatures of 21.2~K and 21.3~K, respectively, indicating a saturation behavior of $T_c$ in the strong-coupling regime. Under biaxial tensile strain, Mo$_2$CI$_2$ exhibits a substantial increase in EPC strength, with $\lambda = 1.60$ at $+1\%$ strain and 1.98 at $+2\%$ strain. However, the accompanying strong phonon softening significantly suppresses the logarithmic average phonon frequency, resulting in moderate superconducting transition temperatures of 15.3~K and 15.7~K, respectively.

% =========================
\section{Conclusions}
% =========================
In summary, we have performed a systematic first-principles study of halogen-functionalized Mo$_2$C MXene monolayers, Mo$_2$C$X_2$ ($X=$ Br, I), focusing on their structural stability, lattice dynamics, electronic properties, and phonon-mediated superconductivity. Our calculations demonstrate that Br- and I-functionalized Mo$_2$C monolayers are dynamically and mechanically stable two-dimensional systems, satisfying the stability criteria for hexagonal lattices and exhibiting robust phonon spectra free of imaginary modes.  

Both monolayers display metallic electronic structures dominated by Mo $d$ states at the Fermi level, providing a favorable electronic environment for strong electron–phonon coupling (EPC). The calculated Eliashberg spectral functions reveal that low- and intermediate-frequency phonon modes associated primarily with Mo vibrations contribute most significantly to the EPC. As a result, Mo$_2$CBr$_2$ and Mo$_2$CI$_2$ are identified as phonon-mediated superconductors in the strong-coupling regime, with superconducting transition temperatures of 13.1 K and 18.1 K, respectively. Importantly, halogen functionalization itself significantly enhances superconductivity in Mo$_2$C. Compared with pristine Mo$_2$C, the introduction of heavier Br and I atoms leads to a substantial increase in the EPC constant, resulting in a pronounced enhancement of the superconducting transition temperature. In particular, I functionalization yields the strongest EPC and the highest $T_c$, highlighting chemical functionalization as an effective route to boosting superconductivity in MXene monolayers.

We further demonstrate that the superconducting properties of these systems are highly tunable under external perturbations. Electron doping substantially enhances the electron–phonon coupling (EPC) strength and leads to a marked increase in the superconducting transition temperature, reaching values above 21 K in both Br- and I-functionalized monolayers. In contrast, biaxial tensile strain induces competing effects: although a substantial enhancement of the EPC constant $\lambda$ is observed, the accompanying strong reduction in the logarithmic average phonon frequency $\omega_{\ln}$ counterbalances this enhancement, resulting in a nearly unchanged $T_c$.

Overall, our results establish halogen-functionalized Mo$_2$C MXenes as promising platforms for two-dimensional superconductivity and highlight carrier doping as an effective strategy for optimizing their superconducting performance. These findings provide valuable insights for the design and experimental realization of tunable superconducting MXene-based materials.

% =========================
% Data Availability
% =========================
    \section*{Data Availability}
    The data that support the findings of this study are available from the corresponding
    authors upon reasonable request.
    
% =========================
% Code Availability
% =========================
    \section*{Code Availability}
    The first-principles DFT calculations were performed using the open-source Quantum ESPRESSO package, available at \url{https://www.quantum-espresso.org}, along with pseudopotentials from the Quantum ESPRESSO pseudopotential library at \url{https://pseudopotentials.quantum-espresso.org/}.
% =========================
% Acknowledgments
% =========================
\section*{Acknowledgments}
	This research project is supported by the Second Century Fund (C2F), Chulalongkorn University. We acknowledge the supporting computing infrastructure provided by NSTDA, CU, CUAASC, NSRF via PMUB [B05F650021, B37G660013] (Thailand).

% =========================
% Author Contributions
% =========================
    \section*{Author Contributions}
    Jakkapat Seeyangnok performed all calculations, analyzed the results, and wrote the initial draft of the manuscript. Udomsilp Pinsook contributed to the analysis and interpretation of the results, and revised the manuscript.
    
% =========================
% References
% =========================
\bibliographystyle{unsrt}
\bibliography{references}

\end{document}